\documentclass[aps,prl,twocolumn,showpacs,preprintnumbers]{revtex4-1}

\usepackage{psfrag,graphicx}
\usepackage{dcolumn}
\usepackage{amsmath,amssymb}
\usepackage{bm}
\usepackage{amsfonts,amssymb,amsmath}        % for math symbols.
\usepackage{epstopdf}
\usepackage{amsthm}
\usepackage{url}

\newcommand{\beq}{\begin{equation}}
\newcommand{\eeq}{\end{equation}}
\newcommand{\bea}{\begin{eqnarray}}
\newcommand{\eea}{\end{eqnarray}}

\newcommand{\ket}[1]{\left | \, #1 \right\rangle}

\begin{document}

\title{Demonstrating non-Abelian braiding of surface code defects in a five qubit experiment}
\author{James R.~Wootton}
\affiliation{Department of Physics, University of Basel, Klingelbergstrasse 82, CH-4056 Basel, Switzerland}

\date{\today}

\begin{abstract}

Currently, the mainstream approach to quantum computing is through surface codes. One way to store and manipulate quantum information with these to create defects in the codes which can be moved and used as if they were particles. Specifically, they simulate the behaviour of exotic particles known as Majoranas, which are a kind of non-Abelian anyon. By exchanging these particles, important gates for quantum computation can be implemented. Here we investigate the simplest possible exchange operation for two surface code Majoranas. This is found to act non-trivially on only five qubits. The system is then truncated to these five qubits, so that the exchange process can be run on the IBM 5Q processor. The results demonstrate the expected effect of the exchange. This paper has been written in a style that should hopefully be accessible to both professional and amateur scientists.

\end{abstract}

\maketitle

Surface codes are the most well known starting point for fault-tolerant quantum computation \cite{dennis:02}. One way to store and manipulate information in these to engineer certain kinds of  defects \cite{bombin:10,brown:16}. These can be moved around and manipulated in much the same way as particles. However, their restriction to the 2D structure of the surface codes allows them to exhibit some of the exotic behaviour possible for particles in 2D universes.

The important features of these particles are their their effects when fused and braided. These refer to the processes of combining particles and of exchanging the positions of particles, respectively. It has been predicted that the fusion rules for these defects identical to so-called Majorana modes or Ising anyons. Their braiding forms a projective representation of the braid group, but is otherwise also identical to that of Majoranas. We will therefore refer to them simply as Majoranas from henceforth.

In this paper we present an experiment performed on a small patch of surface code. The patch is nevertheless large enough for these defects to be introduced and even for pair of them to exchange positions. We implement this and show that the effects of the process are consistent with the braiding of Majoranas, as predicted.

\section{Anyons}

There are many types of particle in the universe, but their properties when exchanged place them into just two categories: bosons and fermions. Neither of which are very complex. This is because a loop around a point in three spatial dimensions can be continuously deformed to a loop that is not around a point. One can simply pick the loop up and place it elsewhere. As such, all topological properties of one particle moving in a loop around another must be trivial. There are only two ways to achieve this: bosonic and fermionic exchange behaviour.

In a two dimensional universe, this no longer holds. There is no longer an `up' to move the loop through, and so it is stuck around the point. The braiding of particles may therefore have non-trivial properties. This would allow almost any type of exchange. Such particles are therefore known collectively as \emph{any}ons \cite{kitaev:06}. For a simple introduction, see \cite{decodoku:anyons}.

Though we do not live in a two-dimensional universe, we can make two dimensional physical systems. This allows us to find phases of matter in which anyons arise as quasi-particles or other localized features of the system than can be moved and otherwise manipulated in the same manner as particles \cite{kitaev:03,kitaev:06}.

\section{Majoranas}

The most interesting families of anyons are the non-Abelian anyons. One example is the Ising anyon model. This holds two types of particle: one known as a Majorana, and the other a fermion denote by $\psi$ \cite{kitaev:06}.

The fermion type $\psi$ is its own antiparticle. Combining two of these will always result in annihilation.

The Majoranas are their own antiparticle, so a pair of them can be created from vacuum. When combined, these would annihilate back to vacuum. However, it is also possible to obtain a pair of Majoranas from the decay of a $\psi$. These would recreate the fermion if brought back together.

These two hypothetical pairs of Majoranas are completely indistinguishable. Their memory of whether to annihilate or form a fermion is not stored in any locally accessible feature. Instead it is stored non-locally through quantum entanglement in the underlying physical system.

This non-locality makes these particles an attractive proposition for storing quantum information in a quantum computer. By associating a pair that annihilate with a bit value $0$, and a pair that form a $\psi$ with bit value $1$, they can be used to store a bit (or qubit). By keeping the Majoranas well separated, it would take a large and concerted effort for errors to affect the bit (by moving the particles together to read out the value, for instance). As such, the information will have an inherent robustness against errors, as long as error correction is performed to detect the errors before they build up \cite{wootton:14,brell:14,hutter:16}.

Now consider two pairs of Majoranas created from vacuum. If one Majorana from one pair is combined with one from the other, there is no shared history to determine the result. The pair will therefore randomly choose to either annihilate or become a $\psi$.

Since everything initially came from vacuum, to vacuum they must return. Combining the remaining pair of Majoranas must then always yield the same result as for the first pair, such than any $\psi$ formed by one will be joined by its antiparticle from the other. It is this process, and the resulting correlation, that is the key to the experiment we will perform.

The exchange of Majoranas, which is simply swapping the positions of two of the anyons, also allows for many interesting effects. For one thing, and as we will show experimentally, it can be used to move particles to be combined with strangers, and so exhibit the effect described above as well as many more complicated versions.

Also, suppose we take again the two pairs of Majoranas created from vacuum. One Majorana from one pair is moved around a Majorana from the other. Combining the two pairs will now result in both forming a $\psi$, and hence flipping the bits that they are storing. This is certainly a non-trivial effect that bosons and fermions could only dream of. It demonstrates the power of non-Abelian anyons to not only store information, but to process it too.

Note that the fault-tolerance described above applies only when the Majoranas can be kept well separated, and when error correction is performed. The small nature of our experimental set-up prevents the possibility of either. As will be seen later, this means that our results will not be free of noise of the effects of noise.

For another approach towards exchanging Majoranas, as well as a summary of all other experimental progress towards Majoranas in a variety of physical systems, see \cite{xu:16,meyer:16} and references therein.

\section{Qubits}

Qubits are the quantum analogue of bits. Just as a bit can be either $0$ or $1$, a qubit can be either $\ket{0}$ or $\ket{1}$. But it can also be any quantum superposition of the two, such as,
$$
\ket{\psi} = a \ket{0} + b \ket{1},
$$
for any arbitrary complex numbers $a$ and $b$ \cite{nielsen:00}. For a simple introduction, see \cite{decodoku:qubits}.

The states $\ket{0}$ and $\ket{1}$ are referred to as the $Z$ basis of the qubit. If we measure to see whether the qubit is $\ket{0}$ or $\ket{1}$, this is called a $Z$  measurement.

Two particular examples of possible superpositions are the states $\ket{+}$ and $\ket{-}$, which are known as the $X$ basis states,
$$
\ket{\pm} = \frac{1}{\sqrt 2}\ket{0} + \frac{\pm 1}{\sqrt 2} \ket{1}.
$$
These are orthogonal states: and as different from each other as $\ket{0}$ is from $\ket{1}$. This means we can also measure whether a qubit is $\ket{+}$ or $\ket{-}$, known as an $X$ measurement. Similarly we may also define so-called $Y$ basis states, and a $Y$ measurement.

The manipulation of qubits can be understood using the circuit model. In this work we will be considering the same circuits as used in the IBM Quantum Experience, and so refer to that for an introduction \cite{IBM:16}.

The most notable operation applied in these circuits is the CNOT. This acts on two qubits, with one called the 'control' and the other the 'target'. The effect is to add the $Z$ basis value of the control to that of the target. However, since the target is restricted to the two states $0$ and $1$, only the parity of the sum is retained. The target is therefore left in the state $0$ if the sum was even, and $1$ if odd. If the target was initially in the $\ket{0}$ state, the effect is to copy the $Z$ basis state of the control to the target. The CNOT gate is a reversible equivalent of the XOR gate used in electronic circuits.

In the IBM 5Q processor there are five qubits, labelled $Q_0$ to $Q_5$. The only CNOTs that can be applied are those with $Q_2$ as the target. In order to allow more flexibility in circuit design, we can implement a CNOT with $Q_2$ as control by applying the Hadamard gate, $H$, on both control and target before and after the standard CNOT is applied.

\section{Matching Code}

\begin{figure}[t]
\begin{center}
{\includegraphics[width=0.7\columnwidth]{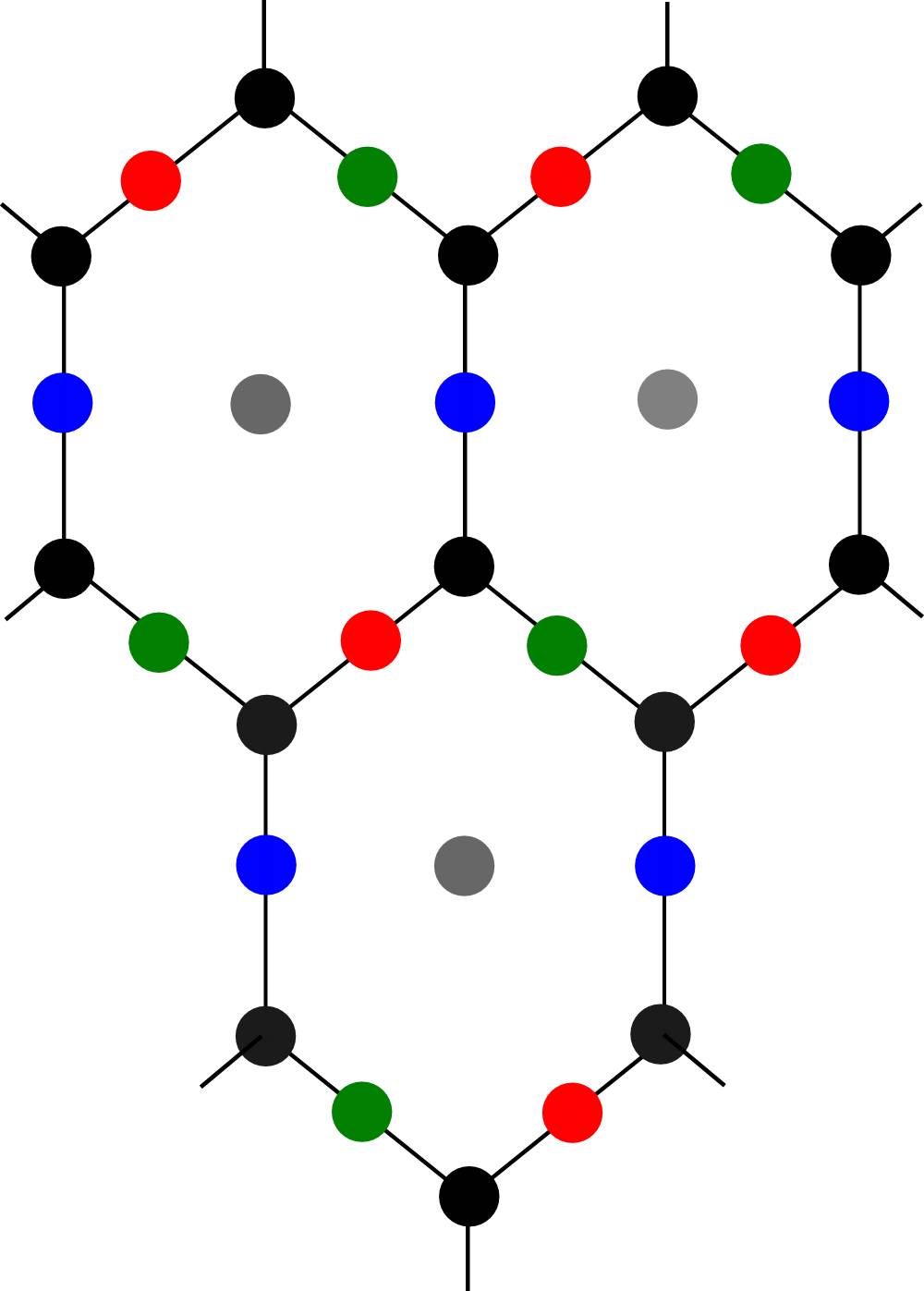}}
\caption{\label{grid} The code is defined on a hexagonal lattice, of which a section is shown. Vertex qubits are shown in black and hexagon in grey. Edge qubits are red, green or blue depending on whether the edge is associated with a $XX$, $YY$ or $ZZ$ measurement, respectively.}
\end{center}
\end{figure}

The grid shown in Fig. \ref{grid} is an example of a matching code \cite{wootton:15}, which is a type of quantum error correcting code \cite{lidar:13} that is based on the honeycomb lattice model \cite{kitaev:06}. It is a variant of the surface code \cite{dennis:02}.

In this code there is a qubit located at each vertex, edge and hexagon. The vertex qubits are those that are truly part of the code. The edge and hexagon qubits are there to make measurements of the vertex qubits.

We associate a measurement with each edge. Each of these measures a collective property of the two qubits on the vertices connected by the edge.

For the vertical edges we have a so-called $ZZ$ measurement. There are two possible outcomes, which correspond to whether or not the state of the qubits different in the $Z$ basis. If both qubits are $\ket{0}$, or both are $\ket{1}$, the measurement outputs $0$ to show there is no difference. The same will happen for a superposition of both being $\ket{0}$ and both being $\ket{1}$, and the measurement will not disturb this superposition. On the other hand, if one qubit is $\ket{0}$ and the other is $\ket{1}$, the measurement will output $1$ to show that there is a difference.

\begin{figure}[t]
\begin{center}
{\includegraphics[width=0.9\columnwidth]{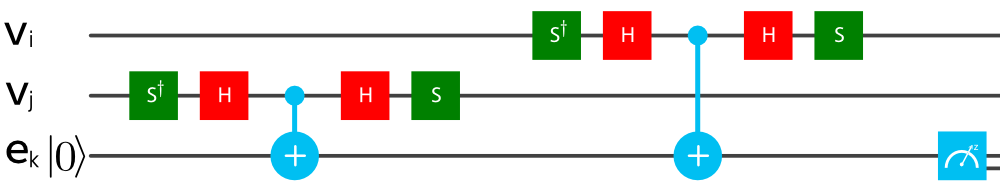}}
\caption{\label{XXYYZZ} In order to implement an $ZZ$ measurement on two vertex qubits $V_i$ and $V_j$ using edge qubit $e_k$, the above circuit is applied using only the blue gates. The first CNOT copies the $Z$ basis state of $v_j$ to $e_k$. The second adds the $Z$ basis state of $v_i$. If their states in the $Z$ basis are the same, the sum is even and so $e_k$ is left in the state $\ket{0}$. If they are different, the sum is odd and so $e_k$ is left in the state $\ket{1}$. The measurement of $e_k$ then outputs the corresponding bit value $0$ or $1$. An $XX$ measurement is the same, except that the red gates are added. These rotate the control end of the CNOTs such that they look at the $X$ basis states, rather than the $Z$. For $YY$ both the red and green gates are added to rotate to the $Y$ basis.}
\end{center}
\end{figure}

The other two edges correspond to $XX$ and $YY$ measurements. These are the same as a $ZZ$ measurement, except that they compares whether the states are the same (output $0$) or different (output $1$) in the $X$ and $Y$ basis, respectively. The circuits that implement these measurements are shown in Fig. \ref{XXYYZZ}.

Measurements that share a vertex qubit do not commute. This means that making a measurement of one will affect the outcome of the other. For example, suppose we prepare a state such that the $ZZ$ measurement on the light blue edge of Fig. \ref{grid} will give the outcome $0$ with certainty. We then measure the $XX$ shown in light red. Subsequently measuring $ZZ$ will then randomly output $0$ or $1$, due to the effect of the non-commuting $XX$. Further explanation of this can be found in \cite{decodoku:measurements}.

The measurements of the hexagons similarly have output either $0$ or $1$. These measure a collective property of all six qubits on the vertices around the hexagon. However, we need not consider these measurements explicitly for what follows. 

All edge measurements commute with all hexagon measurements. We wish to find a set of measurements that all commute with each other, and for which there are as many measurements in the set as there are vertices. The simplest choice is to take all $ZZ$ measurements and all hexagon measurements.

These will form the so-called \emph{stabilizer}, which defines our quantum code. We will prepare and work with states for which the outcome is $0$ to all stabilizer measurements. Measuring any $1$ will then be clue that an error has occurred. However, we will not be considering error correction in what follows.

\section{Anyons in a Quantum Code}

We consider states for which all stabilizer measurements yield the outcome $0$ with certainty. Through the effects of noise, or of our own effort, we can then create states in for which some measurements yield $1$. However, it is impossible to create a single isolated $1$ in some area by solely manipulating qubits in its neighbourhood. Instead, such local manipulations will always cause at least two stabilizers to yield $1$. This is similar to the property of particles, which cannot be created or annihilated in isolation. At least one other particle is required to serve as its antiparticle.

It has been found that stabilizers yielding a $1$ behave as particles in all other ways also. They are therefore quasiparticles that arise from collective properties of the underlying qubits. And these quasiparticles behave as anyons.

The quasiparticles associated with hexagons correspond to the $e$ and $m$ anyons of the surface codes \cite{dennis:02}. However, we will not be concerned with these in what follows.

The quasiparticles residing on the edge stabilizers correspond to fermions. In fact, they behave exactly the same as the $\psi$ fermions of the Ising model. It is therefore interesting to determine whether these can be caused to decay into a pair of Majoranas, and whether these Majoranas can be moved, braided and used as non-Abelian anyons.

This can indeed be done. By changing the definition of the stabilizer close to the $\psi$ we wish to split, we can push the two constituent Majoranas apart. This is shown in Fig. \ref{exchange}, and proven rigorously in \cite{kitaev:06,wootton:15}. This is equivalent to the twist deformations of \cite{bombin:10}.

While performing this process, it is important to keep track of the results of the new stabilizers. Each result of $1$ implies that the Majorana being moved has decayed into a Majorana and a $\psi$ during the process, and so must be recombined. To avoid the need to account for such stray fermions, we will consider only results for which these measurements all yield the result $0$.

\section{Minimal Exchange of Majoranas}

\begin{figure}[t]
\begin{center}
{\includegraphics[width=0.9\columnwidth]{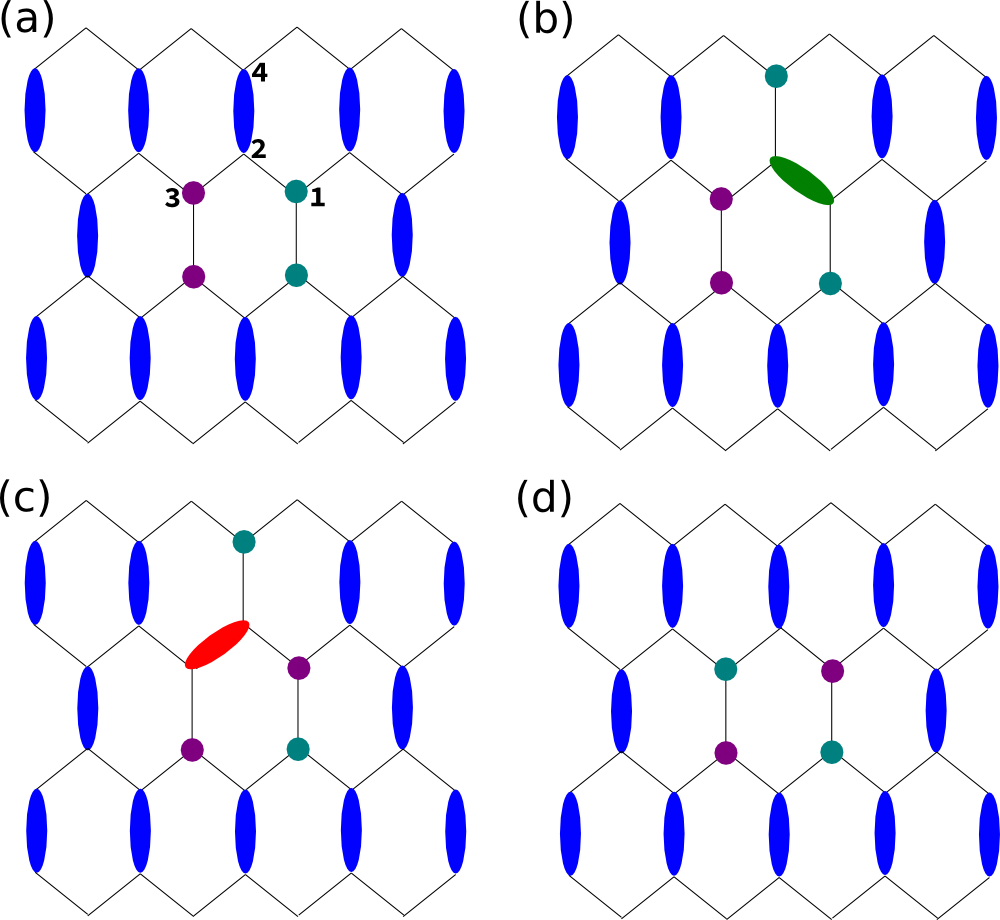}}
\caption{\label{exchange} 
In (a) we see the code with $ZZ$ stabilizers shown in blue. The corresponding pairs of Majoranas can be associated with the vertices. These two pairs are shown explicitly for two $ZZ$ stabilizers shown in purple and teal. (b) To move the green Majorana on vertex $v_1$ we add the adjacent $YY$ measurement to the stabilizer. Since this does not commute with the $ZZ$ measurement above, it is also removed from the stabilizer. The $YY$ measurement binds the Majoranas on vertices $v_1$ and $v_2$ into a well-defined fermionic mode, whereas that on $v_4$ becomes unbound. The Majorana initially at $v_1$ is therefore effectively teleported to $v_4$ \cite{bonderson:08,wootton:15}. (c) An $XX$ measurement is similarly used to teleport the purple Majorana initially at $v_3$ to $v_1$. (d) Finally the initial $ZZ$ measured, and so returned to the stabilizer. This moves the teal Majorana from $v_4$ to $v_3$, completing the exchange.
}
\end{center}
\end{figure}

The process shown in Fig. \ref{exchange} exchanges two Majoranas. The process of this exchange was introduced in \cite{wootton:15}, but is also explained in the caption of the figure.

The implementation of the exchange acts non-trivially only on three vertex qubits and two edge qubits. We may therefore truncate the lattice to this small area, and use a five qubit process to implement the exchange.

\begin{figure}[t]
\begin{center}
{\includegraphics[width=0.7\columnwidth]{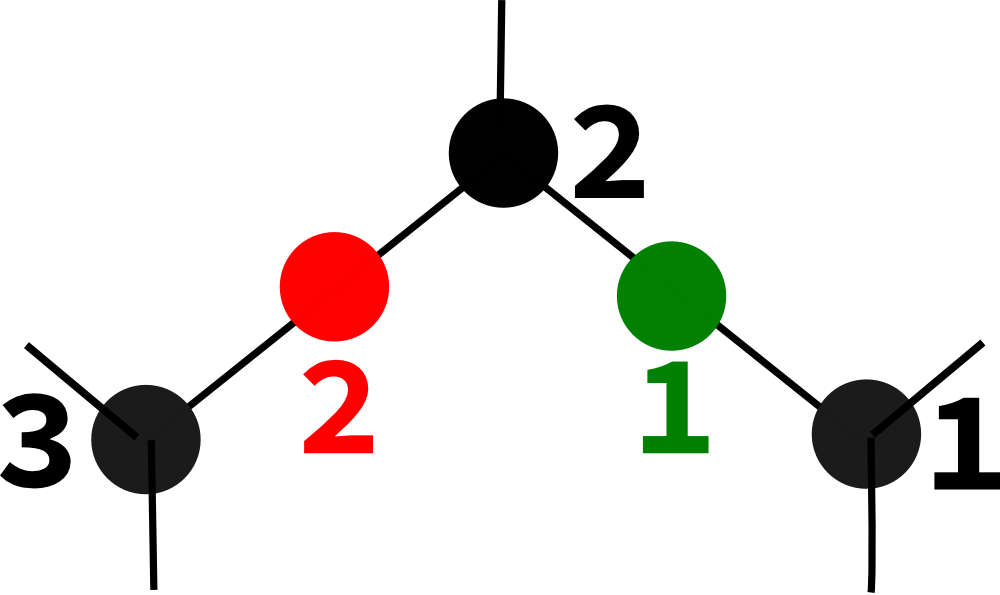}}
\caption{\label{truncated}
The system used for the exchange is reduced to the five qubits on which the process acts non-trivially. The $ZZ$ measurements are reduced to single $Z$ measurements on the vertex qubits. The pairs of Majoranas that residing on these edges therefore now reside fully on the vertices. The $Z$ measurements determine whether they would combine to annihilate or form a fermion, just as the $ZZ$ measurements did previously. Other edge operators remain unchanged.
}
\end{center}
\end{figure}

The truncated system is shown in Fig. \ref{truncated}. In common with other truncated surface code experiments \cite{martinis:14,chow:14,nigg:14}, not all stabilizers have full support on the truncated area. They must therefore be reduced to the part with support on this area. Most notably, the three vertical links incident on the area only have one vertex within it. The $ZZ$ measurements required for the full system then become single $Z$ measurements. The fermionic occupancy of these three links therefore corresponds exactly to the $Z$ basis state of the three vertex qubits. Accordingly, the initial state is simply that for which all vertex qubits are $\ket{0}$. The initial state of the edge qubits, as always, should also be $\ket{0}$.

\begin{figure}[t]
\begin{center}
{\includegraphics[width=0.9\columnwidth]{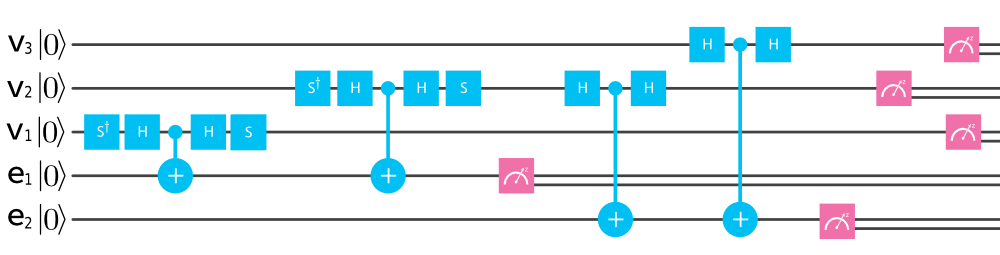}}
\caption{\label{ideal} The ideal circuit first performs the $YY$ measurement of $v_1$ and $v_2$ using $e_1$, and then the $XX$ of $v_2$ and $v_3$ using $e_2$. The $Z$ measurement of $v_2$ is then performed. Afterwards, $Z$ measurements are made on $v_1$ and $v_3$ to verify the expected correlations.}
\end{center}
\end{figure}

The most straightforward implementation of this process is shown in Fig. \ref{ideal}. Here the two edge qubits are used to make the required $XX$ and $YY$ measurements.

This circuit cannot be run on the IBM 5Q processor due to its restricted topology. Because of this we must use a single qubit as the intermediary for both the $XX$ and $ZZ$ measurements. In order for the results of the two measurements to be distinguished, the result of the $YY$ measurement is copied onto an otherwise unused qubit using a CNOT.

\begin{figure}[t]
\begin{center}
{\includegraphics[width=0.9\columnwidth]{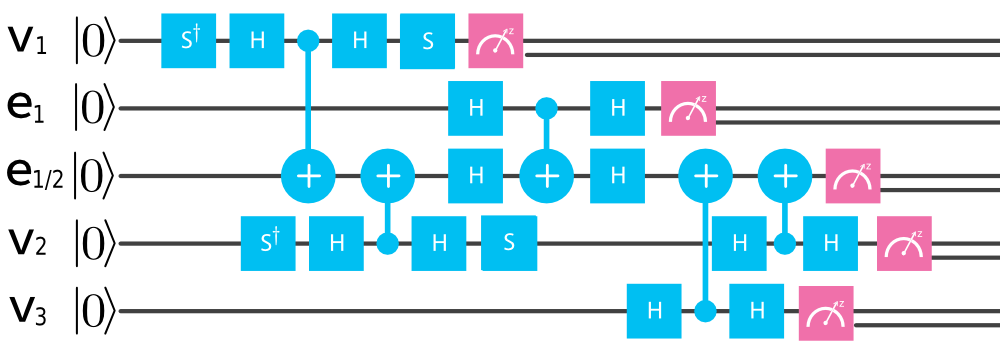}}
\caption{\label{optimized} The $YY$ measurement is made using $e_{1/2}$. But rather than measuring this qubit, the result is copied to $e_1$ for measurement. The $XX$ measurement is then made using $e_{1/2}$. The measurement result of this will be the summ of the $XX$ and $YY$ measurements, but the $YY$ outcome can be inferred using that of $XX$. Each operation is applied as soon as the previous one on the same qubit is complete in order to avoid unnecessary noise. The qubits $Q_0$ to $Q_1$ of the processor are listed from top to bottom. The choice of which should play the role of each $v_j$ and $e_j$ was determined by the $T_1$, $T_2$ and $e^{jk}_g$ parameters from the $2016-09-21$ calibration, in order to minimize noise.}
\end{center}
\end{figure}

It is also important to apply the circuit as quickly as possible, and to assign the qubits to their tasks based on the noise levels of their entangling gates and also their lifetimes. Doing do results in the final circuit as shown in Fig. \ref{optimized}.

\section{Results}

The circuit of Fig. \ref{optimized} was implemented using the IBM 5Q processor, using the interface provided by the IBM quantum experience. The success of the circuit is categorized by the correlation function,
$$
C = P(00) + P(11) - P(01) - P(10).
$$
Here $P(01)$ is the probability that the measurement of $v_1$ yields $0$ and $v_3$ yields $1$, both in the case of no stray fermions. The theoretical prediction for the ideal case is of perfect correlations, corresponding to $C=1$. In contrast $C=0$ would correspond to a lack of correlations and $C=-1$ to perfect anticorrelation. 

A simulation of the circuit under noiseless conditions showed exactly the result expected: $C=1$. A simulation under realistic conditions yields $C=0.454$.

Running the circuit for 24576 shots on the IBM 5Q processor yields $C=0.530$. This is in good agreement with the theoretical prediction made under realistic conditions. It is also well above zero, and so clearly demonstrates the expected correlations.

One possible problem with this result is that decay to the $\ket{0}$ state on $v_1$ and $v_3$ could lead to a false positive. To test this, tomography on the state is performed. Specifically, one can consider the Majoranas to encode a single logical qubit. The exchange has the effect of rotating this qubit to a $Y$ basis state. The logical $X$ operator corresponds to the product of the $XX$ and $YY$ link operators, and so to an $YZX$ operator on $v_1$, $v_2$ and $v_3$. Respectively. Since the logical $Z$ is simply a single $Z$ on $v_1$ or $v_3$, the $Y$ will be $XZX$ or $YZY$ respectively.

Since these three qubit measurements are performed at readout, superpositions in the system need not be preserved. We may therefore do three single qubit measurements and calculate the corresponding correlators afterwards. This simply requires additional rotations on $v_1$ and $v_3$ of Fig \ref{optimized} to access the correct basis. The results for each are obtained from 8192 shots.

With these results, we can reconstruct the density matrix of the resulting state and compute its fidelity to the required $Y$ basis state. It is found that this fidelity is $70.6\%$ when using $YZY$. The closest point on the surface of the Bloch sphere has a fidelity of $94.0\%$ to the required state. This is therefore a very encouraging result.

When using $XZX$ the state obtained has a fidelity of only $54.6\%$, and that of the closest point on the Bloch sphere is $67.4\%$. The poor result in this case is most likely due to the longer time needed to measure the qubit $v_1=Q_0$, which has the lowest lifetime of all the qubits.

All data can be found at \cite{wootton:results}.

\section{Conclusion}

The exchange of two Majoranas causes the state of two Majorana pairs with definite fusion result to become one with indefinite but correlated results. In this work we showed that this could be realized on a surface code based architecture using only five qubits. The experiment was then performed using the IBM 5Q processor. The results obtained demonstrate the expected correlations.

With higher fidelity qubits, this work could be extended by applying two exchanges. The effect of a full braid of one Majorana around another could then be demonstrated, as well as the effect of undoing the first exchange by one in the opposite direction. The orthogonality of the results in these two cases will provide an even starker demonstration of the braiding of non-Abelian anyons.

It would also be interesting to see this experiment reproduced with spin qubits. This could use the process proposed in \cite{engel:05} or the system introduced in \cite{veldhorst:15}.

\section{Acknowledgements}

We acknowledge use of the IBM Quantum Experience for this work. The views expressed are those of the author and do not reflect the official policy or position of IBM or the IBM Quantum Experience team. The author was supported by the NCCR QSIT.


\begin{thebibliography}{22}
\expandafter\ifx\csname natexlab\endcsname\relax\def\natexlab#1{#1}\fi
\expandafter\ifx\csname bibnamefont\endcsname\relax
  \def\bibnamefont#1{#1}\fi
\expandafter\ifx\csname bibfnamefont\endcsname\relax
  \def\bibfnamefont#1{#1}\fi
\expandafter\ifx\csname citenamefont\endcsname\relax
  \def\citenamefont#1{#1}\fi
\expandafter\ifx\csname url\endcsname\relax
  \def\url#1{\texttt{#1}}\fi
\expandafter\ifx\csname urlprefix\endcsname\relax\def\urlprefix{URL }\fi
\providecommand{\bibinfo}[2]{#2}
\providecommand{\eprint}[2][]{\url{#2}}

\bibitem[{\citenamefont{Dennis et~al.}(2002)\citenamefont{Dennis, Kitaev,
  Landahl, and Preskill}}]{dennis:02}
\bibinfo{author}{\bibfnamefont{E.}~\bibnamefont{Dennis}},
  \bibinfo{author}{\bibfnamefont{A.}~\bibnamefont{Kitaev}},
  \bibinfo{author}{\bibfnamefont{A.}~\bibnamefont{Landahl}}, \bibnamefont{and}
  \bibinfo{author}{\bibfnamefont{J.}~\bibnamefont{Preskill}},
  \bibinfo{journal}{J. Math. Phys.} \textbf{\bibinfo{volume}{43}},
  \bibinfo{pages}{4452} (\bibinfo{year}{2002}).

\bibitem[{\citenamefont{Bombin}(2010)}]{bombin:10}
\bibinfo{author}{\bibfnamefont{H.}~\bibnamefont{Bombin}},
  \bibinfo{journal}{Phys. Rev. Lett.} \textbf{\bibinfo{volume}{105}},
  \bibinfo{pages}{030403} (\bibinfo{year}{2010}).

\bibitem[{\citenamefont{Brown et~al.}(2016)\citenamefont{Brown, Laubscher,
  Kesselring, and Wootton}}]{brown:16}
\bibinfo{author}{\bibfnamefont{B.~J.} \bibnamefont{Brown}},
  \bibinfo{author}{\bibfnamefont{K.}~\bibnamefont{Laubscher}},
  \bibinfo{author}{\bibfnamefont{M.~S.} \bibnamefont{Kesselring}},
  \bibnamefont{and} \bibinfo{author}{\bibfnamefont{J.~R.}
  \bibnamefont{Wootton}} (\bibinfo{year}{2016}),
  \bibinfo{note}{arXiv:1609.04673}.

\bibitem[{\citenamefont{Kitaev}(2006)}]{kitaev:06}
\bibinfo{author}{\bibfnamefont{A.}~\bibnamefont{Kitaev}},
  \bibinfo{journal}{Ann. Phys.} \textbf{\bibinfo{volume}{321}},
  \bibinfo{pages}{2} (\bibinfo{year}{2006}).

\bibitem[{\citenamefont{Wootton}(2016{\natexlab{a}})}]{decodoku:anyons}
\bibinfo{author}{\bibfnamefont{J.~R.} \bibnamefont{Wootton}},
  \emph{\bibinfo{title}{Anyons}} (\bibinfo{year}{2016}{\natexlab{a}}),
  \urlprefix\url{http://decodoku.blogspot.ch/2016/06/anyons.html}.

\bibitem[{\citenamefont{Kitaev}(2003)}]{kitaev:03}
\bibinfo{author}{\bibfnamefont{A.~Y.} \bibnamefont{Kitaev}},
  \bibinfo{journal}{Ann. Phys.} \textbf{\bibinfo{volume}{303}},
  \bibinfo{pages}{2} (\bibinfo{year}{2003}).

\bibitem[{\citenamefont{Wootton et~al.}(2014)\citenamefont{Wootton, Burri,
  Iblisdir, and Loss}}]{wootton:14}
\bibinfo{author}{\bibfnamefont{J.~R.} \bibnamefont{Wootton}},
  \bibinfo{author}{\bibfnamefont{J.}~\bibnamefont{Burri}},
  \bibinfo{author}{\bibfnamefont{S.}~\bibnamefont{Iblisdir}}, \bibnamefont{and}
  \bibinfo{author}{\bibfnamefont{D.}~\bibnamefont{Loss}},
  \bibinfo{journal}{Phys. Rev. X} \textbf{\bibinfo{volume}{4}},
  \bibinfo{pages}{011051} (\bibinfo{year}{2014}).

\bibitem[{\citenamefont{Brell et~al.}(2014)\citenamefont{Brell, Burton,
  Dauphinais, Flammia, and Poulin}}]{brell:14}
\bibinfo{author}{\bibfnamefont{C.~G.} \bibnamefont{Brell}},
  \bibinfo{author}{\bibfnamefont{S.}~\bibnamefont{Burton}},
  \bibinfo{author}{\bibfnamefont{G.}~\bibnamefont{Dauphinais}},
  \bibinfo{author}{\bibfnamefont{S.~T.} \bibnamefont{Flammia}},
  \bibnamefont{and} \bibinfo{author}{\bibfnamefont{D.}~\bibnamefont{Poulin}},
  \bibinfo{journal}{Phys. Rev. X} \textbf{\bibinfo{volume}{4}},
  \bibinfo{pages}{031058} (\bibinfo{year}{2014}).

\bibitem[{\citenamefont{Hutter and Wootton}(2016)}]{hutter:16}
\bibinfo{author}{\bibfnamefont{A.}~\bibnamefont{Hutter}} \bibnamefont{and}
  \bibinfo{author}{\bibfnamefont{J.~R.} \bibnamefont{Wootton}},
  \bibinfo{journal}{Phys. Rev. A} \textbf{\bibinfo{volume}{93}},
  \bibinfo{pages}{042327} (\bibinfo{year}{2016}).

\bibitem[{\citenamefont{Xu et~al.}(2016)\citenamefont{Xu, Sun,
  Han, Li, Pachos and Guo}}]{xu:16}
\bibinfo{author}{\bibfnamefont{J.-S.}~\bibnamefont{Xu}},
  \bibinfo{author}{\bibfnamefont{K.}~\bibnamefont{Sun}},
  \bibinfo{author}{\bibfnamefont{Y.-J.}~\bibnamefont{Han}},
  \bibinfo{author}{\bibfnamefont{C.-F.}~\bibnamefont{Li}},
  \bibinfo{author}{\bibfnamefont{J.K.}~\bibnamefont{Pachos}}, \bibnamefont{and}
  \bibinfo{author}{\bibfnamefont{G.-C.}~\bibnamefont{Guo}},
  \bibinfo{journal}{Nat. Comms.} \textbf{\bibinfo{volume}{7}},
  \bibinfo{pages}{13194} (\bibinfo{year}{2016}).

\bibitem[{\citenamefont{Pawlak et~al.}(2016)\citenamefont{Pawlak, Kisiel, Klinovaja, Meier, Kawai, Glatzel, Loss and Meyer}}]{meyer:16}
\bibinfo{author}{\bibfnamefont{R.}~\bibnamefont{Pawlak}},
  \bibinfo{author}{\bibfnamefont{M.}~\bibnamefont{Kisiel}},
  \bibinfo{author}{\bibfnamefont{J.}~\bibnamefont{Klinovaja}},
  \bibinfo{author}{\bibfnamefont{T.}~\bibnamefont{Meier}},
\bibinfo{author}{\bibfnamefont{S.}~\bibnamefont{Kawai}},
\bibinfo{author}{\bibfnamefont{T.}~\bibnamefont{Glatzel}},
  \bibinfo{author}{\bibfnamefont{D.}~\bibnamefont{Loss}}, \bibnamefont{and}
  \bibinfo{author}{\bibfnamefont{E.}~\bibnamefont{Meyer}},
  \bibinfo{journal}{npj Quant. Inf.} \textbf{\bibinfo{volume}{2}},
  \bibinfo{pages}{16035} (\bibinfo{year}{2016}).

\bibitem[{\citenamefont{Nielsen and Chuang}(2000)}]{nielsen:00}
\bibinfo{author}{\bibfnamefont{M.~A.} \bibnamefont{Nielsen}} \bibnamefont{and}
  \bibinfo{author}{\bibfnamefont{I.~L.} \bibnamefont{Chuang}},
  \emph{\bibinfo{title}{Quantum Computation and Quantum Information}}
  (\bibinfo{publisher}{{Cambridge University Press}},
  \bibinfo{address}{Cambridge}, \bibinfo{year}{2000}).

\bibitem[{\citenamefont{Wootton}(2016{\natexlab{b}})}]{decodoku:qubits}
\bibinfo{author}{\bibfnamefont{J.~R.} \bibnamefont{Wootton}},
  \emph{\bibinfo{title}{Qubits with the simplest maths possible}}
  (\bibinfo{year}{2016}{\natexlab{b}}),
  \urlprefix\url{http://decodoku.blogspot.ch/2016/02/the-maths-of-qubits.html}.

\bibitem[{\citenamefont{IBM}(2016)}]{IBM:16}
\bibinfo{author}{\bibnamefont{IBM}}, \emph{\bibinfo{title}{Quantum experience}}
  (\bibinfo{year}{2016}),
  \urlprefix\url{https://quantumexperience.ng.bluemix.net/qstage/#/community}.

\bibitem[{\citenamefont{Wootton}(2015)}]{wootton:15}
\bibinfo{author}{\bibfnamefont{J.~R.} \bibnamefont{Wootton}},
  \bibinfo{journal}{Journal of Physics A: Mathematical and Theoretical}
  \textbf{\bibinfo{volume}{48}}, \bibinfo{pages}{215302}
  (\bibinfo{year}{2015}).

\bibitem[{\citenamefont{Lidar and Brun}(2013)}]{lidar:13}
\bibinfo{editor}{\bibfnamefont{D.~A.} \bibnamefont{Lidar}} \bibnamefont{and}
  \bibinfo{editor}{\bibfnamefont{T.~A.} \bibnamefont{Brun}}, eds.,
  \emph{\bibinfo{title}{{Quantum Error Correction}}}
  (\bibinfo{publisher}{Cambridge University Press},
  \bibinfo{address}{{Cambride, UK}}, \bibinfo{year}{2013}).

\bibitem[{\citenamefont{Wootton}(2016{\natexlab{c}})}]{decodoku:measurements}
\bibinfo{author}{\bibfnamefont{J.~R.} \bibnamefont{Wootton}},
  \emph{\bibinfo{title}{Fun measurements for pairs of qubits}}
  (\bibinfo{year}{2016}{\natexlab{c}}),
  \urlprefix\url{http://decodoku.blogspot.ch/2016/04/fun-measurements-for-pairs-of-qubits.html}.

\bibitem[{\citenamefont{Bonderson et~al.}(2008)\citenamefont{Bonderson,
  Freedman, and Nayak}}]{bonderson:08}
\bibinfo{author}{\bibfnamefont{P.}~\bibnamefont{Bonderson}},
  \bibinfo{author}{\bibfnamefont{M.}~\bibnamefont{Freedman}}, \bibnamefont{and}
  \bibinfo{author}{\bibfnamefont{C.}~\bibnamefont{Nayak}},
  \bibinfo{journal}{Phys. Rev. Lett.} \textbf{\bibinfo{volume}{101}},
  \bibinfo{pages}{010501} (\bibinfo{year}{2008}).

\bibitem[{\citenamefont{Kelly~{\em et al.}}(2014)}]{martinis:14}
\bibinfo{author}{\bibfnamefont{J.}~\bibnamefont{Kelly~{\em et al.}}},
  \bibinfo{journal}{Nature} \textbf{\bibinfo{volume}{519}}, \bibinfo{pages}{66}
  (\bibinfo{year}{2014}).

\bibitem[{\citenamefont{Corcoles~{\em et al.}}(2015)}]{chow:14}
\bibinfo{author}{\bibfnamefont{A.~D.} \bibnamefont{Corcoles~{\em et al.}}},
  \bibinfo{journal}{Nature Communications} \textbf{\bibinfo{volume}{6}}
  (\bibinfo{year}{2015}).

\bibitem[{\citenamefont{Nigg et~al.}(2014)\citenamefont{Nigg, M{\"u}ller,
  Martinez, Schindler, Hennrich, Monz, Martin-Delgado, and Blatt}}]{nigg:14}
\bibinfo{author}{\bibfnamefont{D.}~\bibnamefont{Nigg}},
  \bibinfo{author}{\bibfnamefont{M.}~\bibnamefont{M{\"u}ller}},
  \bibinfo{author}{\bibfnamefont{E.~A.} \bibnamefont{Martinez}},
  \bibinfo{author}{\bibfnamefont{P.}~\bibnamefont{Schindler}},
  \bibinfo{author}{\bibfnamefont{M.}~\bibnamefont{Hennrich}},
  \bibinfo{author}{\bibfnamefont{T.}~\bibnamefont{Monz}},
  \bibinfo{author}{\bibfnamefont{M.~A.} \bibnamefont{Martin-Delgado}},
  \bibnamefont{and} \bibinfo{author}{\bibfnamefont{R.}~\bibnamefont{Blatt}},
  \bibinfo{journal}{Science} \textbf{\bibinfo{volume}{345}},
  \bibinfo{pages}{302} (\bibinfo{year}{2014}).

\bibitem[{\citenamefont{Wootton}(2016{\natexlab{d}})}]{wootton:results}
\bibinfo{author}{\bibfnamefont{J.~R.} \bibnamefont{Wootton}},
  \emph{\bibinfo{title}{Raw data}} (\bibinfo{year}{2016}{\natexlab{d}}),
  \urlprefix\url{https://www.dropbox.com/sh/2o50l9tcl5l0962/AADUv46t_NooxkrnqzdXH2jva?dl=0}.

\bibitem[{\citenamefont{Engel and Loss}(2005)}]{engel:05}
\bibinfo{author}{\bibfnamefont{H.-A.} \bibnamefont{Engel}} \bibnamefont{and}
  \bibinfo{author}{\bibfnamefont{D.}~\bibnamefont{Loss}},
  \bibinfo{journal}{Science} \textbf{\bibinfo{volume}{309}},
  \bibinfo{pages}{586} (\bibinfo{year}{2005}), ISSN \bibinfo{issn}{0036-8075}.

\bibitem[{\citenamefont{Veldhorst et~al.}(2015)\citenamefont{Veldhorst, Yang,
  Hwang, Huang, Dehollain, Muhonen, Simmons, Laucht, Hudson, Itoh
  et~al.}}]{veldhorst:15}
\bibinfo{author}{\bibfnamefont{M.}~\bibnamefont{Veldhorst}},
  \bibinfo{author}{\bibfnamefont{C.~H.} \bibnamefont{Yang}},
  \bibinfo{author}{\bibfnamefont{J.~C.~C.} \bibnamefont{Hwang}},
  \bibinfo{author}{\bibfnamefont{W.}~\bibnamefont{Huang}},
  \bibinfo{author}{\bibfnamefont{J.~P.} \bibnamefont{Dehollain}},
  \bibinfo{author}{\bibfnamefont{J.~T.} \bibnamefont{Muhonen}},
  \bibinfo{author}{\bibfnamefont{S.}~\bibnamefont{Simmons}},
  \bibinfo{author}{\bibfnamefont{A.}~\bibnamefont{Laucht}},
  \bibinfo{author}{\bibfnamefont{F.~E.} \bibnamefont{Hudson}},
  \bibinfo{author}{\bibfnamefont{K.~M.} \bibnamefont{Itoh}},
  \bibnamefont{et~al.}, \bibinfo{journal}{Nature}
  \textbf{\bibinfo{volume}{526}}, \bibinfo{pages}{410} (\bibinfo{year}{2015}),
  ISSN \bibinfo{issn}{0028-0836}.

\end{thebibliography}
\end{document}